\documentclass[aps,amsfonts,epsf]{revtex4}
\usepackage{graphics}
\usepackage{graphicx}
\usepackage{epsfig}

\def\beq{\begin{equation}}
\def\eeq{\end{equation}}
\def\bea{\begin{eqnarray}}
\def\eea{\end{eqnarray}}

\begin{document}
%% \normalsize  
\input epsf.tex

\title{The Effect of Higher Harmonic Corrections on the Detection of massive black hole binaries with LISA.}
\author{Edward K. Porter$^{1}$ and Neil J. Cornish$^{2}$} 
\affiliation{$^{1}$Max Planck Institut f\"ur Gravitationsphysik, Albert Einstein Institut, Am M\"uhlenberg 1, Golm bei Potsdam, D-14476, Germany.\\
$^{2}$Department of Physics, Montana State University, Bozeman, 59717, MT, USA.}
\vspace{1cm}

\vspace{1cm}
\begin{abstract}
Massive black hole binaries are key targets for the space based gravitational wave interferometer LISA.
Several studies have investigated how LISA observations could be used to constrain the parameters of these
systems. Until recently, most of these studies have ignored the higher harmonic corrections to the waveforms.
Here we analyze the effects of the higher harmonics in more detail by performing extensive Monte Carlo simulations.
We pay particular attention to how the higher harmonics impact parameter correlations, and show that the additional
harmonics help mitigate the impact of having two laser links fail, by allowing for an instantaneous measurement of
the gravitational wave polarization with a single interferometer channel. By looking at parameter correlations we are able
to explain why certain mass ratios provide dramatic improvements in certain parameter estimations, and illustrate
how the improved polarization measurement improves the prospects for single interferometer operation.
\\

\end{abstract}

\maketitle
%%%%%%%%%%%%%%%%%%%%%%%%%%%%%%%%%%%%%%%%%%%%%%%%%%%%%%%%%%%%%%%%%%%%%%%%%%%%%%%%
%%%%%%%%%%%%%%%%%%%%%%%%%%%%%
%%%%%%%%%%%%%%%%%%%%%%%%%%%%%%%%%%% Introduction 
%%%%%%%%%%%%%%%%%%%%%%%%%%%%%%%%%%%%%%%%%%%%%%%%%%%%%%%%%%%%
%%%%%%%%%%%%%%%%%%%%%%%%%%%%%%%%%%%%%%%%%%%%%%%%%%%%%%%%%%%%%%%%%%%%%%%%%%%%%%%%
%%%%%%%%%%%%%%%%%%%%%%%%%%%%%

\section{Introduction}
Massive black hole binaries are expected to be some of the brightest gravitational wave (GW) sources for the planned Laser Interferometer Space Antenna (LISA)~\cite{lisa}.  This joint ESA-NASA mission will search for GWs is the band $10^{-5}\leq f/{\rm Hz} \leq 1$.  While the
event rates are uncertain, it is likely that there will be at least a few events per year.  Massive black hole binaries are a very promising
source of GWs as we expect them to have integrated signal to noise ratios (SNRs) in the hundreds to thousands.  As well as being very
bright, there should be little confusion with other sources (a problem which plagues galactic binary extraction, and to a lesser extent
the extreme mass ratio inspiral sources).

There have been many studies looking at parameter estimation for spinning and non-spinning binaries~\cite{cc, sv, mh1, mh2,
sh1, bbw, av, lh1, lh2}, and, more recently, on the development of search algorithms~\cite{en1, en2, jpl, lisafrance, stas}.
With the exception of Refs.~\cite{sv,mh1,mh2}, these studies have focused on the contribution from the dominant second harmonic
of the orbital frequency, and have neglected the influence of the additional harmonics that appear at higher post-Newtonian order.
The studies that did consider the impact of these higher harmonic corrections (HHCs) saw improvements in parameter estimation,
but only a few cases were considered. More extensive studies of the impact of HHCs have appeared in the last year~\cite{ts,arun1, arun2},
and it has become apparent the effects can vary significantly from source to source. 

Our goal here is to consider a wider sample of systems, and to elucidate the mechanism by which the HHCs affect
parameter estimation. Our most significant finding is that HHCs greatly mitigate the effects of any hardware
failures that reduce the instrument to a single interferometry channel~\cite{noord}. Indeed, when HHCs are included,
the science performance with a single channel is comparable to what was found for two channels when HHCs are
neglected. 

We performed extensive Monte Carlo simulation for various redshifted chirp masses, with coalescence time $t_c = 1$,
lumiosity distance $D_L = 10\,{\rm Gpc}$ and all other parameters chosen at random. Results for other distances can be
obtained by multiplying our parameter uncertainities by $10\,{\rm Gpc}/D_L$. As sky resolution is an important quantity,
we considered 20,000 different sky locations per chirp mass and per mass ratio.  Finally, for each chirp mass, we run
separate simulations for mass ratios of 1 and 10.

\subsection{Outline of the paper}
The paper is structured as follows.  In Sec~(\ref{sec:gw}) we outline the form of the gravitational waveform at the detector with higher harmonics corrections.  We also give a brief description of the low frequency approximation~\cite{cc} for the LISA response.  Sec.~(\ref{sec:pe}) contains a brief outline of the main tools used for parameter estimation.  We finish this section with an outline of the Monte Carlo simulations we carried out.  In Sec~(\ref{sec:hhc}) we present results of our analysis on the effect of the higher harmonic corrections for parameter estimation using different redshifted chirp masses and mass ratios.  This is followed in Sec~(\ref{sec:cb}) by an investigation into what correlation breaking between parameters is responsible for the increase in parameter estimation.  The final main section, Sec~(\ref{sec:ovt}), deals with the effect of losing interferometry links during the mission.  

\section{The Gravitational Waveform}\label{sec:gw}
In the low frequency approximation, we can think of LISA as being composed of two orthogonal $90^o$ detectors.  The strain of the gravitational wave (GW) in each channel of the LISA detector with both polarizations is given by
\begin{equation}
h(t) = h_{+}(\xi(t))F^{+}+h_{\times}(\xi(t))F^{\times}, 
\end{equation}
where the phase shifted time parameter is 
\begin{equation}
\xi(t) = t - R_{\oplus}\sin\theta\cos\left(\alpha(t) - \phi\right).
\end{equation}
Here, $R_{\oplus} = 1 AU \approx$ 500 secs is the radial distance to the detector guiding center, $\left(\theta,\phi\right)$ are the position angles of the source in the sky, $\alpha(t)=2\pi f_{m}t + \kappa$, $f_{m}=1/year$ is the LISA modulation frequency and $\kappa$ gives the initial ecliptic longitude of the guiding center.  The GW polarizations up to 2-PN order in amplitude corrections are defined by~\cite{biww}
\begin{equation}
h_{+,\times}= \frac{2Gm\eta}{c^{2}D_{L}}x\,\left[H_{+,\times}^{(0)}+x^{1/2}H_{+,\times}^{(1/2)}+xH_{+,\times}^{(1)}+x^{3/2}H_{+,\times}^{(3/2)}+x^{2}H_{+,\times}^{(2)}\right].
\label{eqn:strain}
\end{equation}
Here $m=m_{1}+m_{2}$ is the total mass of the binary, $\eta = m_{1}m_{2}/m^{2}$ is the reduced mass ratio and $D_{L}$ is the luminosity distance of the source.  The invariant PN velocity parameter is defined by, $x = \left(Gm\omega / c^{3}\right)^{2/3}$, where $\omega=d\Phi_{0rb}/dt$ is the 2 PN order orbital frequency for a circular orbit and $\Phi_{orb}=\varphi_{c}^{orb}-\phi_{orb}(t)$ is the orbital phase which is defined as
\begin{equation}
\Phi_{orb}(t) = \varphi_{c}^{orb}-\frac{1}{\eta}\left\{\Theta^{5/8}+\left(\frac{3715}{8064}+\frac{55}{96}\eta\right)\Theta^{3/8}-\frac{3\pi}{4}\Theta^{1/4} +\left(\frac{9275495}{14450688}+\frac{284875}{258048}\eta+\frac{1855}{2048}\eta^{2}\right)\Theta^{1/8}\right\},
\label{eqn:phase}
\end{equation}
where the quantity $\Theta$ is related to the time to coalescence of the wave, $t_{c}$, by
\begin{equation}
\Theta(t) = \frac{c^{3}\eta}{5Gm}\left(t_{c}-t\right),
\end{equation}
and $\varphi_{c}^{orb}$ is the orbital phase of the wave at coalescence.  All GW phases are then twice the orbital value.  For the rest of the paper, we will work with GW phases.

In Eqn~(\ref{eqn:strain}), the functions $H_{+,\times}^{(n)}$ contain the PN corrections to the amplitude and the extra phase harmonics.  The restricted PN waveform corresponds to keeping just the $H_{+,\times}^{(0)}$ terms.  We should note here that contained in the half integer $H_{+,\times}^{(n)}$ terms is a factor $\delta m = m_1 - m_2$.  This term has the effect of killing all the odd phase harmonics in the equal mass case.  We can also see the extra frequency harmonics arising due to the $x^n$ terms.

Using the WMAP values of $(\Omega_{R}, \Omega_{M}, \Omega_{\Lambda}) = (4.9\times10^{-5}, 0.27, 0.73)$ and a Hubble's constant of $H_{0}$=71 km/s/Mpc, the relation between redshift, $z$, and luminosity distance, $D_{L}$, is given by
\begin{equation}
D_{L} = \frac{c(1+z)}{H_{0}}\int_{0}^{z}dz'\left[\Omega_{R}\left(1+z'\right)^{4}+\Omega_{M}\left(1+z'\right)^{3} + \Omega_{\Lambda}\right]^{-1/2}  .
\end{equation}

The functions $F^{+,\times}$ are the beam pattern functions of the detector given in the low frequency approximation by 
\beq
F^{+}(t;\psi, \theta, \phi, \lambda) = \frac{1}{2}\left[\cos(2\psi)D^{+}(t;\theta, \phi, \lambda) - \sin(2\psi)D^{\times}(t;\theta, \phi, \lambda)\right],
\eeq
\beq
F^{\times}(t;\psi, \theta, \phi, \lambda) = \frac{1}{2}\left[\sin(2\psi)D^{+}(t;\theta, \phi, \lambda) + \cos(2\psi)D^{\times}(t;\theta, \phi, \lambda)\right],
\eeq
where $\psi$ is the polarization angle of the wave and $\lambda = 0$ or $\pi/4$ defines the two-arm combination of LISA from which the strain is coming.  The detector pattern functions are given by~\cite{cornishrubbo}
\bea
D^{+}(t) &=& \frac{\sqrt{3}}{64}\left[\frac{}{}-36\sin^{2}(\theta)\sin(2\alpha(t)-2\lambda)+(3+\cos(2\theta)) \left(\frac{}{}\cos(2\phi)\left\{\frac{}{}9\sin(2\lambda)-\sin(4\alpha(t)-2\lambda)\right\} \frac{}{}
\right.\right. \\ \nonumber
&+&\left.\left.\sin(2\phi)\left\{\frac{}{}\cos\left(4\alpha(t)-2\lambda\right)-9\cos(2\lambda) \right\}\frac{}{}\right)-4\sqrt{3}\sin(2\theta)\left(\frac{}{}\sin(3\alpha(t)-2\lambda-\phi)\right.\right.\\
&-&\left.\left.3\sin(\alpha(t)-2\lambda+\phi)\frac{}{}\right)\frac{}{}\right] \nonumber,
\eea
\bea
D^{\times}(t) = \frac{1}{16}\left[\frac{}{}\sqrt{3}\cos(\theta)\left(\frac{}{}9\cos(2\lambda-2\phi)-\cos(4\alpha(t)-2\lambda-2\phi) \right) \right. \\ \nonumber \left. -6\sin(\theta)\left(\frac{}{} \cos(3\alpha(t)-2\lambda-\phi)+3\cos(\alpha(t)-2\lambda+\phi) \right) \right].
\eea
For two Schwarzschild black holes, the above equations governing the evolution of the phase break down even before we reach the last stable circular orbit (LSO) at $R=6M$.  Because of this, we terminate the waveforms at $R=7M$.

The low frequency approximation is an extremely good fit to the full detector response at frequencies of $\lesssim$ 3 mHz~\cite{cornishrubbo}.  In this case the two channel formalism originally derived by Cutler~\cite{cc} corresponds to the construction of optimal orthogonal time delay interferometry (TDI) variables $\{A,E\}$ using the unequal-arm Michelson TDI variables $\{X,Y,Z\}$ according to 
\begin{equation}
A = X\,\,\,\,\,\,\, ,\,\,\,\,\,\, E = (X+2Y) / \sqrt{3}.
\end{equation}
In later sections we will refer to the one channel case as $X$ and the two channel as $AE$.

\section{Estimating parameter errors using the Fisher matrix.}\label{sec:pe}
One of the main tools used in the GW community for the estimation of parameter errors is the Fisher information matrix (FIM).  In the high SNR limit, the inverse of the FIM gives the variance-covariance matrix.  The square root of the diagonal elements of the inverse FIM give a 1-$\sigma$ estimation of the error in our parameter estimation.  For LISA, it has been shown in a number of cases that the FIM is a good parameter error estimator when compared with Markov Chain Monte Carlo techniques.  For this particular problem, the parameter set we chose to work with is $\lambda^{\mu} = \{\ln(M_{c}), \ln(\mu), \ln(t_{c}), \cos\theta, \phi, \ln(D_{L}), \cos\iota, \varphi_{c}, \psi\}$ where $M_{c}=m\eta^{3/5}$ is the chirp mass, $\mu =m\eta$ is the reduced mass and all other parameters have been previously defined.  It has been customary to divide this parameter set into intrinsic (parameters that effect the dynamics of the system) and extrinsic parameters (those that are more relative to the detector).  In the parameter set defined above, we consider the final four parameters to be extrinsic.  The Fisher matrix is thus defined by
\begin{equation}
\Gamma_{ij}=\left\langle \frac{\partial h}{\partial\lambda^{i}} \left|\frac{\partial h}{\partial\lambda^{j}}\right\rangle\right.  ,
\end{equation}
where $h = \tilde{h}(f)$ is an un-normalized template and the angular brackets denote the inner product
\begin{equation}\label{eqn:scalarprod}
\left<h\left|s\right.\right> =2\int_{0}^{\infty}\frac{df}{S_{n}(f)}\,\left[ \tilde{h}(f)\tilde{s}^{*}(f) +  \tilde{h}^{*}(f)\tilde{s}(f) \right],
\label{eq:scalarprod}
\end{equation}
where a tilde denotes a Fourier transform and an asterisk denotes a complex conjugate.  The quantity $S_{n}(f)=S_{n}^{instr}(f)+S_{n}^{conf}(f)$ is the one-sided noise spectral density of the detector, which is a combination of instrumental and galactic confusion noise.  For the instrumental noise we use the expression given by~\cite{cornish}
\begin{equation}
S_{n}^{instr}(f)=\frac{1}{4L^{2}}\left[ 2 S_{n}^{pos}(f)\left(2+\cos^{2}\left(\frac{f}{f_{*}}\right)\right)+8 S_{n}^{acc}(f)\left(1+\cos^{2}\left(\frac{f}{f_{*}}\right) \right)\left(\frac{1}{(2\pi f)^{4}}+\frac{\left(2\pi 10^{-4}\right)^2}{(2\pi f)^{6}}\right)\right] , 
\end{equation}
where $L=5\times10^{6}$ km is the arm-length for LISA,  $S_{n}^{pos}(f) = 4\times10^{-22}\,m^{2}/Hz$ and $S_{n}^{acc}(f) = 9\times10^{-30}\,m^{2}/s^{4}/Hz$ are the position and acceleration noise respectively.  The quantity $f_{*}=1/(2\pi L)$ is the mean transfer frequency for the LISA arm.  Notice that the final term in the expression has the effect of reddening the noise below $10^{-4}$ Hz to account for the fact that we may not be able to achieve the desired noise spectral density as we approach $10^{-5}$ Hz. For the galactic confusion we use the following confusion noise estimate derived from a Nelemans, Yungelson, Zwart (NYZ) galactic foreground model~\cite{NYZ, TRC}
\begin{equation}
S_{n}^{conf}(f) = \left\{ \begin{array}{ll} 10^{-44.62}f^{-2.3} & 10^{-4} < f\leq 10^{-3} \\ \\ 10^{-50.92}f^{-4.4} & 10^{-3} < f\leq 10^{-2.7}\\ \\ 10^{-62.8}f^{-8.8} &  10^{-2.7} < f\leq 10^{-2.4}\\ \\ 10^{-89.68}f^{-20} &  10^{-2.4} < f\leq 10^{-2}  \end{array}\right.,
\end{equation}
where the confusion noise has units of $Hz^{-1}$.  When two channels are available the total FIM is the sum of the FIM in each
channel, i.e. $\Gamma_{ij}=\Gamma_{ij}^{I}+\Gamma_{ij}^{II}$.

For this study the Fisher matrix is calculated numerically, where the derivatives of the waveforms are calculated using the central difference equation
\begin{equation}
\frac{\partial h}{\partial\lambda^{\nu}} = \frac{h(\lambda^{\nu}+\Delta\lambda^{\nu}) - h(\lambda^{\nu}-\Delta\lambda^{\nu})}{2\Delta\lambda^{\nu}},
\end{equation}
for all parameters other than $M_c$ and $\mu$ in the equal mass case.  The central difference equation in un-applicable for these two parameters when we have equal masses, as it is a degenerate space.  Certain shifts in one parameter, while holding the other constant, leads to un-astrophysical individual masses.  Therefore, in the equal mass case, for $M_c$ and $\mu$ we use
\begin{equation}
\frac{\partial h}{\partial M_c} = \frac{h(M_c+\Delta M_c) - h(M_c)}{\Delta M_c},
\end{equation}
and
\begin{equation}
\frac{\partial h}{\partial \mu} = \frac{h(\mu) - h(\mu-\Delta\mu)}{\Delta\mu}.
\end{equation}
For all other cases, we revert to the central difference formula.

One of the quantities that we are most interested in investigating is the error box on the sky for a particular source.  We therefore define the positional error on the sky as 
\begin{equation}
 \Delta\Omega = 2\pi \sqrt{\Sigma^{\theta\theta}\Sigma^{\phi\phi}-\left(\Sigma^{\theta\phi}\right)^{2}},
\end{equation}
where the elements of the variance-covariance matrix are given by
\begin{equation}
 \Sigma^{ij} = \left<\Delta\lambda^{i}\Delta\lambda^{j}\right> = \left(\Gamma^{ij}\right)^{-1},
\end{equation}
and the three main quantities in the sky resolution expression are given by
\begin{equation}
 \Sigma^{\theta\theta} = \left<\Delta\cos\theta\Delta\cos\theta\right>,
\end{equation}
\begin{equation}
 \Sigma^{\phi\phi} = \left<\Delta\phi\Delta\phi\right>,
\end{equation}
and
\begin{equation}
 \Sigma^{\theta\phi} = \left<\Delta\cos\theta\Delta\phi\right>.
\end{equation}

Finally, we also quote the optimal signal to noise ratio (SNR) for each source.  In each individual detector this is defined by
\begin{equation}
\rho_{i}^{2} = \left<h_{i}\left|h_{i}\right.\right>.
\end{equation}
When we use both LISA detectors the total SNR is given by
\begin{equation}
\rho = \sqrt{\rho_{I}^2 + \rho_{II}^2},
\end{equation}

For the Monte Carlo simulation, we used $2\times10^{4}$ points, where the redshifted chirp mass, mass ratio, time to coalescence and luminosity distance were held constant and all other parameters were varied.  To investigate as many cases as possible, we chose chirp masses of $M_{c}(z)= \{10^{8}, 10^{7}, 10^{6}, 10^{5},10^{4}\}\,M_{\odot} $.  For each chirp mass, we ran a Monte Carlo for mass ratios of 1 and 10.  In all cases, the sources were put at a constant distance of 10 Gpc.  We assume the time of observation is 1 year, and the time to coalescence in each case is 0.999 years.  To be consistent, we chose not to evolve our templates beyond 5 mHz as we know that we can not trust the low frequency approximation beyond this value. For cases where coalescence is seen (i.e. the coalescence frequency of the maximum harmonic is less than 5 mHz) the templates were terminated once the distance between the bodies reached $7 M$.

\section{The effect of higher harmonic corrections on parameter estimation.}\label{sec:hhc}
For clarity, we will treat the equal and unequal mass results separately.  In both cases, we focus on the most interesting parameters from an astronomical point of view.  We focus on the two mass parameters, the time of coalescence, the luminosity distance and the sky resolution.  For completeness, we also include information on the signal to noise ratios.  We should also point out that due to the large tails in the
distributions, median values are more informative than mean values.   

\begin{figure}[t]
%\vspace{0.25 in}
\begin{center}
\centerline{\epsfxsize=12cm \epsfysize=8.5cm \epsfbox{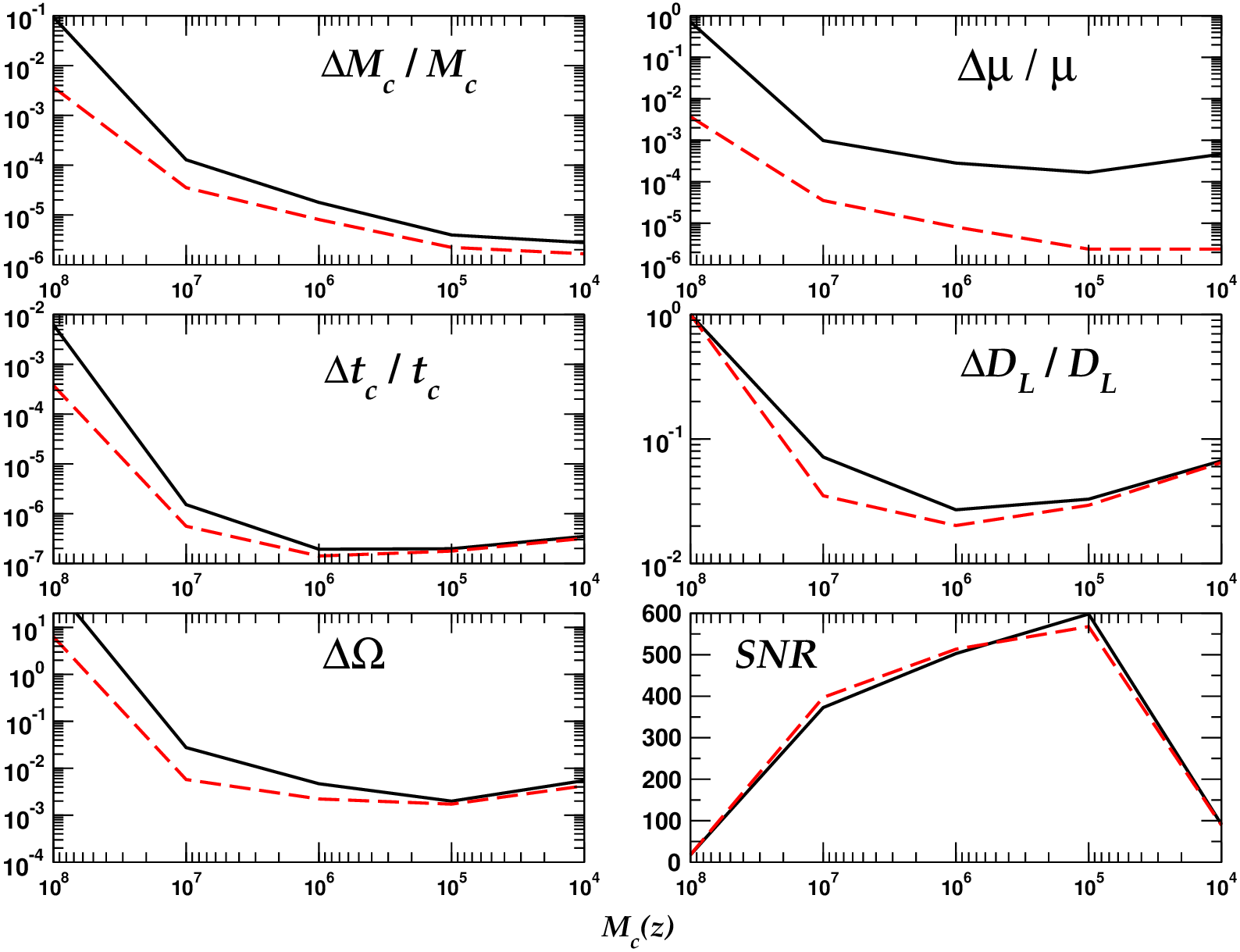}}
%\vspace{2mm}
\caption{A comparison of parameter extraction for equal mass binaries using restricted (solid line) and higher harmonic corrected (dashed line) waveforms, as a function of redshifted chirp mass and at a distance of 10 Gpc.  The values quoted are median values from a $2\times10^4$ point Monte Carlo simulation.} 
\label{fig:em1}
\end{center}
\end{figure}

\begin{figure}[!h]
%\vspace{0.25 in}
\begin{center}
\centerline{\epsfxsize=12cm \epsfysize=8.5cm \epsfbox{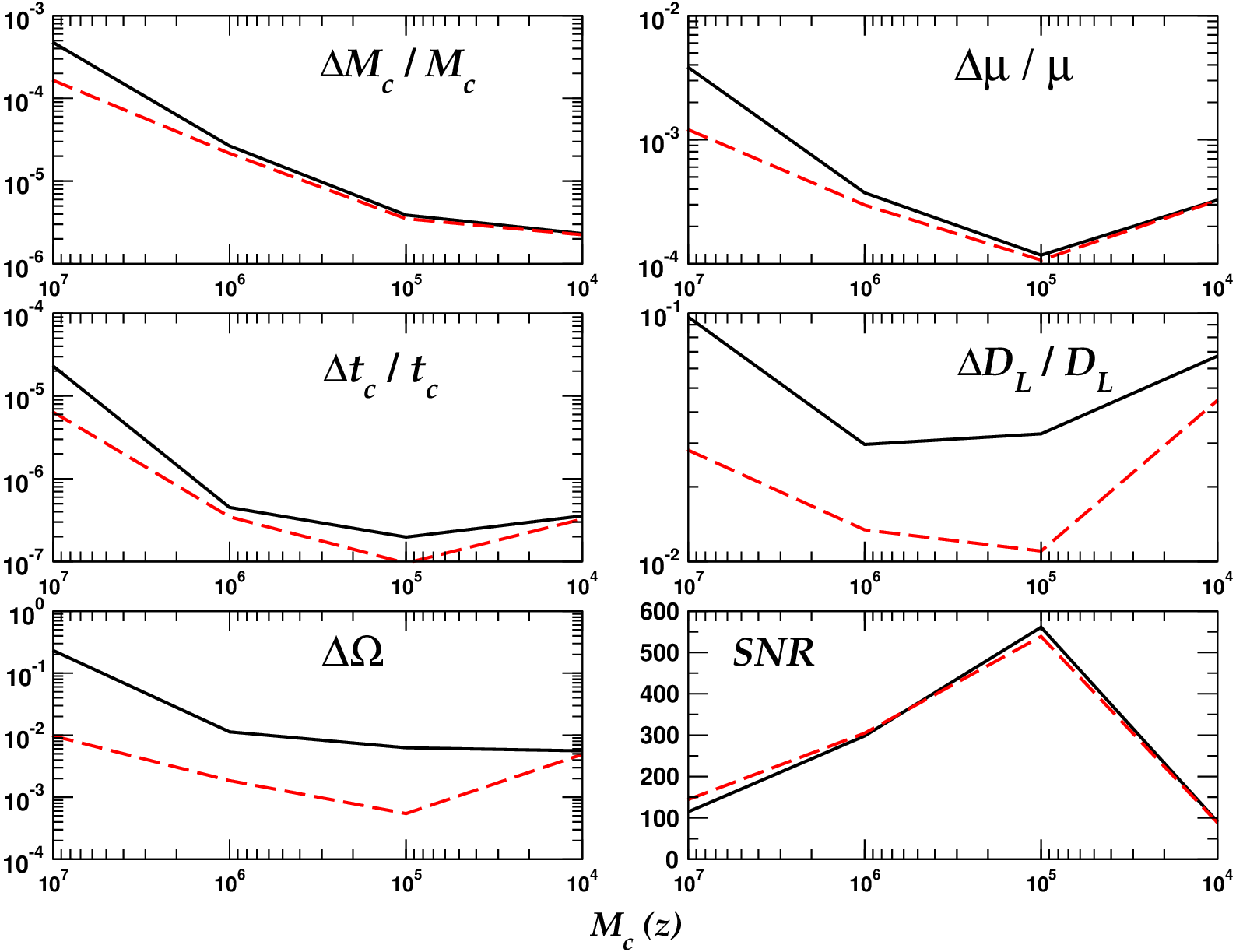}}
%\vspace{2mm}
\caption{A comparison of parameter extraction for binaries with a mass ratio of 10 using restricted (solid line) and higher harmonic corrected (dashed line) waveforms, as a function of redshifted chirp mass and at a distance of 10 Gpc.  The values quoted are median values from a $2\times10^4$ point Monte Carlo simulation.}
\label{fig:uem1}
\end{center}
\end{figure}

\subsection{Equal mass binaries.}
In Fig~(\ref{fig:em1}) we plot the median results from the Monte Carlo simulation, for equal mass binaries, as a function of redshifted chirp mass.  We can see that at 10 Gpcs, we should be able to see sources with redshifted chirp masses of $10^8\,M_{\odot}$, both with and without the extra corrections.  The HHCs give median improvements of factors of 26 in chirp mass, 194 in reduced mass, 17 in time to coalescence and 12 in sky resolution.  However, we should adapt a cautionary air about these sources.  While they are detectable, it does not look like we will be say much about parameter estimation.  Two of the most important quantities for astronomical observations are sky position and distance.  For these particular sources, the distance to the source is unresolvable. The factor of 12 improvement in sky resolution minimizes the error from all-sky to 1 steradian.  Again we should point out that these are median values, so there will be a small subset of the results which will give bigger improvements in the position and distance.  But even then, it may not be good enough for astronomical purposes.

So it is more realistic in this scenario to focus on the lower mass systems for parameter estimation.  In this case we obtain more modest results, except in the case of the reduced mass.  From the Monte Carlo we obtain HHC improvements in chirp mass of about 3.5 at $10^7\,M_{\odot}$ down to approximately 1.7 at $10^4\,M_{\odot}$.  For coalescence time we obtain improvements of 2.7 at high mass to 1 at low mass and an improvement in sky resolution of 5.3 at high mass to 1.3 at low mass.  The smallest improvement comes for luminosity distance with 1.3 at  $10^7\,M_{\odot}$ to 1 at $10^4\,M_{\odot}$.  However, the biggest improvement is for the reduced mass.  Here we get a factor of 28 improvement at high mass, increasing to a factor of 73 at low mass.  We will explain this fantastic result in the next section when we look at correlation breaking between the parameters.  It is clear from the plot that the biggest gain from using the HHC waveforms is in the redshifted chirp mass range of $10^7 \leq M_c/M_{\odot}\leq 5\times 10^5$.  Below this value, the higher harmonics are now out of band at high frequencies.  This leaves only the $H_{+,\times}^{(1/2)}$ correction in band.  From the cell where we display SNR, we can see that this first correction has the effect of subtracting SNR.  The result that the HHC waveform has greater SNR at high mass, and lower SNR at low mass is well known.

\subsection{Unequal mass binaries.}
In Fig~(\ref{fig:uem1}) we plot the same values for the mass ratio 10 scenario.  In this case we are not able to detect binaries with a redshifted chirp mass of $10^8\,M_{\odot}$, even with the inclusion of HHCs.  For unequal mass binaries, the effect of the HHCs  is somewhat less dramatic for the mass parameters.  In fact we can see that we only have improvements in parameter estimation above $10^6\,M_{\odot}$.  Once again, between $10^7\,M_{\odot}$ and $10^4\,M_{\odot}$, we have improvement factors of approximately 3 to 1 for both chirp mass and reduced mass.  However, for $t_{c}, \Delta\Omega$ and $D_{L}$, there is an obvious advantage to having the HHCs.  While the improvement is not great ($4\sim1$), we do see a gain in the time to coalescence estimate.  The improvements for luminosity distance are between 3.5 and 1.5, but the greatest improvement is for the sky resolution with median factors of 24 at $10^7\,M_{\odot}$ and 1.5 at $10^4\,M_{\odot}$.  What is important in this case is that for both of these parameters, the improvement stretches to pretty much all lower masses.  In terms of SNR, we see the same pattern as before.  At higher masses the HHC waveforms have a slightly higher SNR, whereas the opposite is true at low mass.

We should say a few words here about why the HHCs are having an effect.  This was outlined in Refs~\cite{mh1,mh2}, but we feel it is useful to reiterate it here.  We have seen that our measured signal is a function of nine parameters.  The three most important, and hence easiest to measure, parameters are $\{M_c,\mu,t_c\}$.  We can see from Eqn~(\ref{eqn:phase}) that the waveform phase, and from $\omega=d\Phi/dt$,  that the frequency and all its derivatives are determined with these three parameters.  To get good a good estimate for sky position we rely on two effects : first, the Doppler shift which is a function of $\{\theta,\phi\}$ and the beam pattern functions which are a function of $\{\theta,\phi, \psi\}$.  For a system like the inspiral of a massive black hole binary, in effect each detector measures an amplitude and a phase.  Because the waveforms are long-lived in the detector, we are able to measure the frequency and it's derivatives to high accuracy.  This allows us to measure $\{M_c,\mu,t_c\}$ quite accurately, but still leaves six unknowns $\{\theta,\phi, \psi, \iota, \varphi_c, \ln(D_L)\}$.

In a two detector system with no HHCs, each detector measures a phase and an amplitude, giving in effect, four observables with six unknowns.  However, the introduction of the HHCs rectifies this.  In the equal mass case, we stated that all harmonics which are a function of $\delta m=m_1 - m_2$ are null.  This means that if we include all harmonic corrections up to $H^{(2)}_{+,\times}$, we still retain three harmonics, giving us three phases and three amplitudes in each detector.  This gives us twelve observables in total for six unknowns.  For the unequal mass case, we now end up with twenty-four observables for six unknowns.  We can see the effect of having the extra observables by noting that the parameter estimation is always better in the unequal mass case.

\section{Correlation breaking due to the corrected waveforms.}\label{sec:cb}
In the previous section we demonstrated that the HHCs lead to smaller predicted errors in the estimation of parameters, especially in the higher mass range.  While we would expect as improvement due to the fact that the HHC waveforms extend to higher frequencies, this still does not answer the question of what exactly is causing the improvement.  To try and give a more definitive reason, we look at correlation breaking between parameters.  

\begin{figure}[t]
\vspace{0.25 in}
\begin{center}
\centerline{\epsfxsize=14cm \epsfysize=12cm \epsfbox{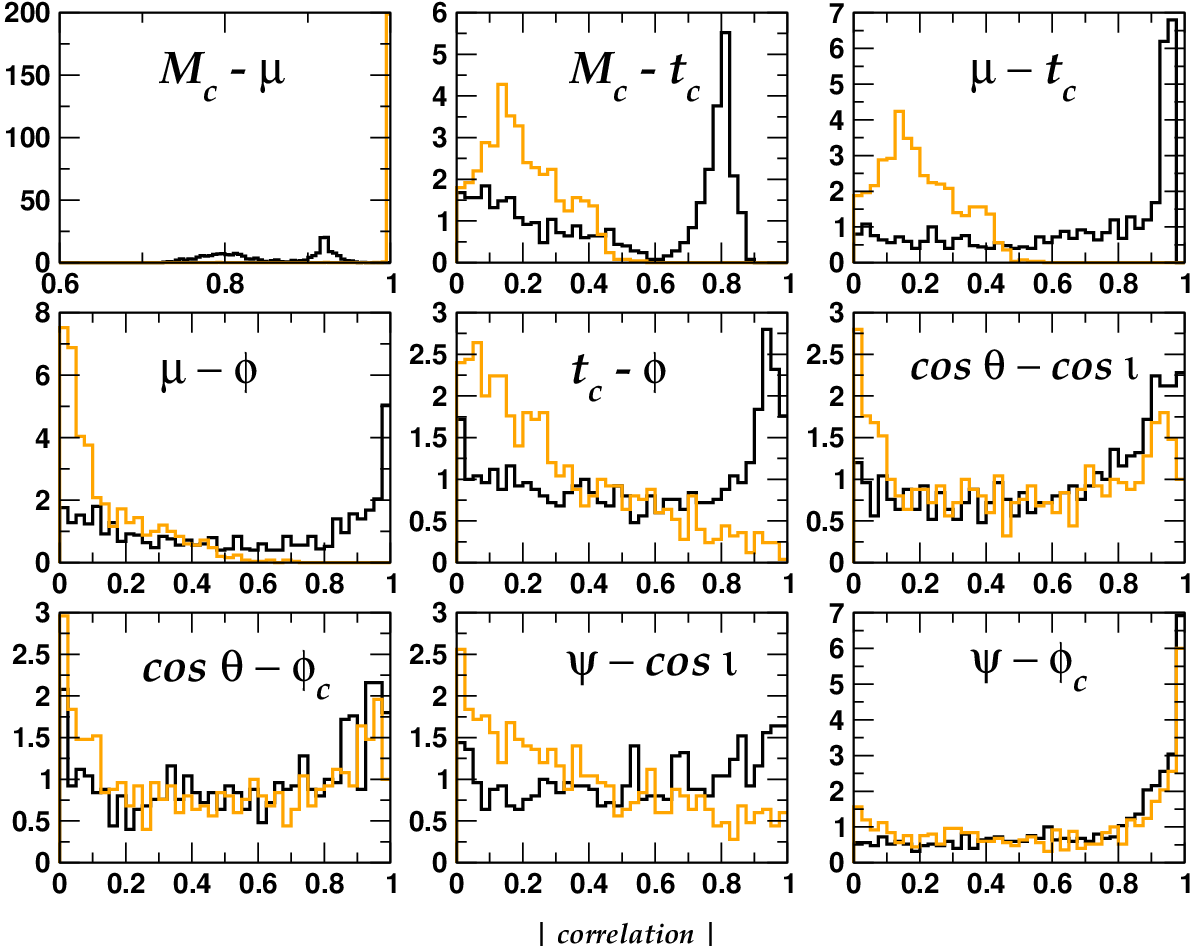}}
\vspace{5mm}
\caption{Main correlation breaking for equal mass binaries with a chirp mass of $10^7\,M_{\odot}$ due to the inclusion of higher harmonic corrections.  The restricted waveform correlations are given by the dark line and the HHC correlations are given by the lighter line.}
\label{fig:emcb}
\end{center}
\end{figure}
\vspace{2mm}

We stated earlier that the inverse of the FIM, $\Gamma_{ij}$, is the variance-covariance matrix $\Sigma^{ij}$.  Rather than working with this, it is more informative to work with the matrix of standard deviations and correlation coefficients defined by
\begin{equation}
 D^{ij} = \left\{ \begin{array}{ll} \sqrt{\Sigma^{ij}} & i=j \\ \\ \Sigma^{ij}/\sqrt{\Sigma^{ii}\Sigma^{jj}} & i \neq j \end{array}\right. .
\end{equation}
In this matrix, the diagonal elements range between $[-1,1]$, where 1 is perfect correlation, -1 is perfect anticorrelation and 0 represents no correlation.  For supermassive black holes there is some degree of correlation between most of the parameters.  However, in most cases the correlations are mild.  As we are concerned with any correlation breaking between the strongly correlated parameters, we will focus on parameter correlations with an absolute value of $>0.5$.  As we have seen that the HHCs have a greater effect on the higher mass binaries, for this exercise we will only concern ourselves with the $M_c=10^7\,M_{\odot}$ case.  Again, we will treat the equal and unequal mass cases separately.

\begin{figure}[t]
\vspace{0.25 in}
\begin{center}
\centerline{\epsfxsize=14cm \epsfysize=12cm \epsfbox{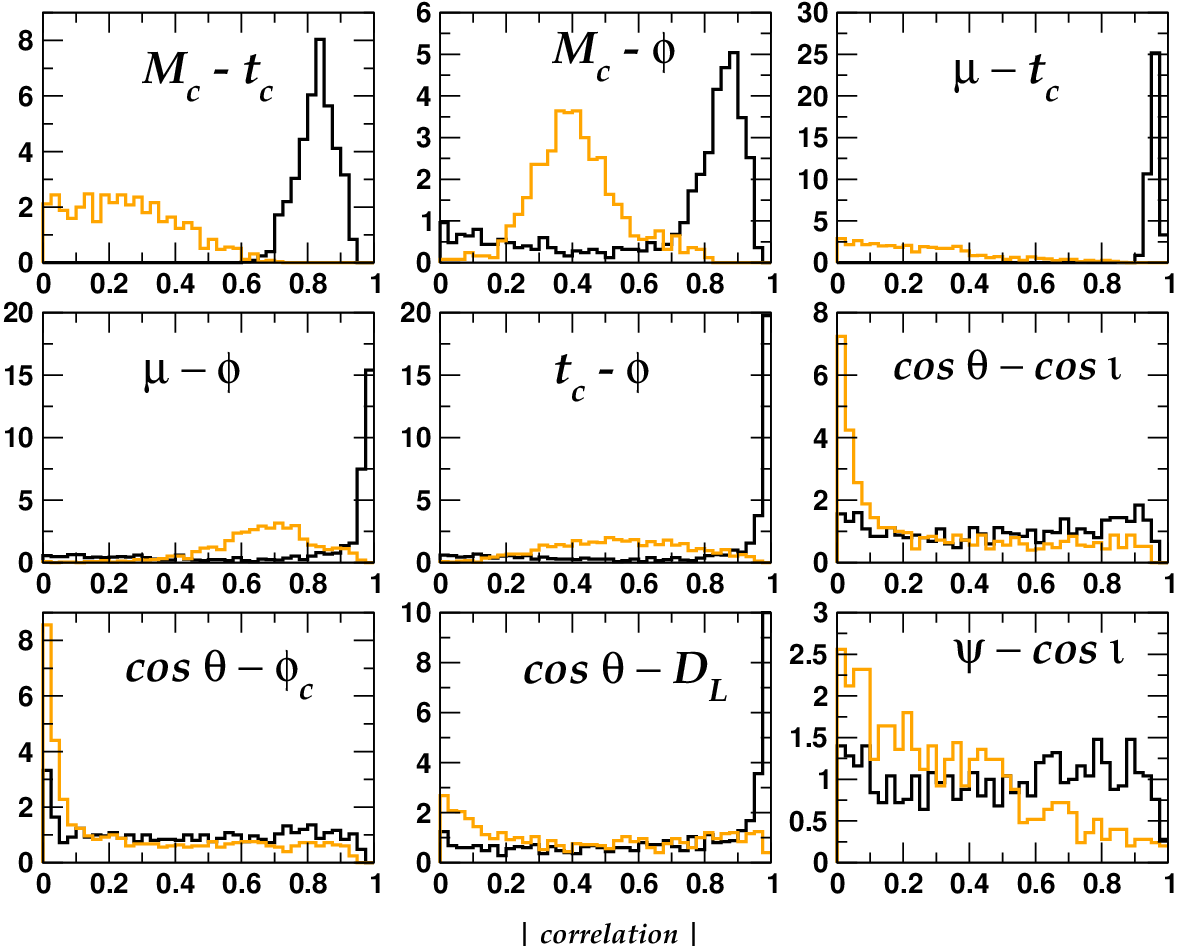}}
\vspace{5mm}
\caption{Main correlation breaking for unequal mass binaries  with a chirp mass of $10^7\,M_{\odot}$ due to the inclusion of higher harmonic corrections.  The restricted waveform correlations are given by the dark line and the HHC correlations are given by the lighter line.}
\label{fig:umcb}
\end{center}
\end{figure}

\subsection{Equal Mass Binaries.}
In Fig~(\ref{fig:emcb}) we plot the strongly correlated parameter breaking for the case of equal mass binaries.  The dark lines represent the restricted waveforms, while the light lines denote the HHC waveforms.  As was shown in Ref~\cite{en1, en2, en3}, there is a strong correlation between $\{M_{c},\mu,t_{c}\}$ in the restricted PN case.  What was not clear in these previous works is the fact that these three parameters are also quite highly correlated with the azimuthal sky angle $\phi$.  On the other hand, it turns out that $\cos\theta$ is highly correlated with the three extrinsic parameters $\{\cos\iota,D_{L}, \varphi_c\}$.  Finally in the restricted case, one of the other parameters of interest is the polarization angle $\psi$ which is correlated mostly with $\{\cos\iota, \varphi_c\}$

If we now focus on the light curves, we can see the effect of introducing the higher harmonic.  The most surprising thing we can immediately see is that for $\{M_{c},\mu\}$, rather than breaking the correlation, the HHCs actual make the reduced mass perfectly correlated with the chirp mass.  In fact, the Fisher elements for these two parameters are numerically equivalent in the equal mass case.  As the chirp mass is already well resolved, even in the restricted waveform case, this perfect correlation explains the huge increase in precision for the reduced mass in Fig~(\ref{fig:em1}).  Going back to the top two cells in this figure, we can see that the error curves for both parameters are almost identical.

So while there is an increase in correlation for $\{M_{c},\mu\}$, we can see that there is a huge correlation breaking for all the other intrinsic parameters upon introduction of the HHCs.  We can see that both $M_{c}$ and $\mu$ decouple from $t_c$, with median absolute correlations dropping from (0.46, 0.78) to (0.18, 0.18) respectively.  In terms of the improvement in sky resolution, it looks like there is a combined effect from the higher harmonics.

The first is that the azimuthal sky parameter $\phi$ decouples from $\{t_{c},\mu\}$ with correlations dropping from (0.55, 0.56) to (0.24, 0.08) respectively.  Secondly, the polar angle $\cos\theta$ decouples from $\{\cos\iota,\varphi_c\}$.  Here the correlations decrease from (0.66, 0.58) to (0.49, 0.48).  Finally we see the same thing happening with the polarization angle $\psi$ with correlation decreasing from (0.55, 0.81) to (0.32, 0.66) respectively.  We believe these various breaking of correlation between $\{\cos\theta,\phi,\psi\}$ and the other parameters are the main reason for the increase in sky resolution.

\subsection{Unequal Mass Binaries.}
In Fig~(\ref{fig:umcb}) we plot the same thing for the unequal mass case. Here we find a similar story to the equal mass case.  The main difference is that in this case we do observe some slight correlation breaking between $M_c$ and $\mu$ with the correlation reducing from 0.94 to 0.87.  Once again, the main intrinsic correlation is between $\{M_{c},\mu,t_{c},\phi\}$.  Firstly, both $M_c$ and $\mu$ decouple from $t_c$ with correlations dropping from (0.83, 0.96) to (0.24, 0.22) respectively.  This is accompanied by correlation breaking between $\{M_{c},\mu,t_{c}\}$ and $\phi$ with respective correlations of (0.83, 0.97, 0.98) dropping to (0.4, 0.68, 0.54).

Again we see a drop in correlation between $\cos\theta$ and $\{\cos\iota, D_L,\varphi_c\}$ with values changing from (0.53, 0.8, 0.46) to (0.19, 0.42, 0.19).  And finally, as in the equal mass case, we also observe some correlation breaking between $\psi$ and $\{\cos\iota,\varphi_c\}$ with correlations being reduced from (0.52, 0.79) to (0.28, 0.68) respectively.

\begin{figure}[t]
\vspace{0.25 in}
\begin{center}
\centerline{\epsfxsize=14cm \epsfysize=10cm \epsfbox{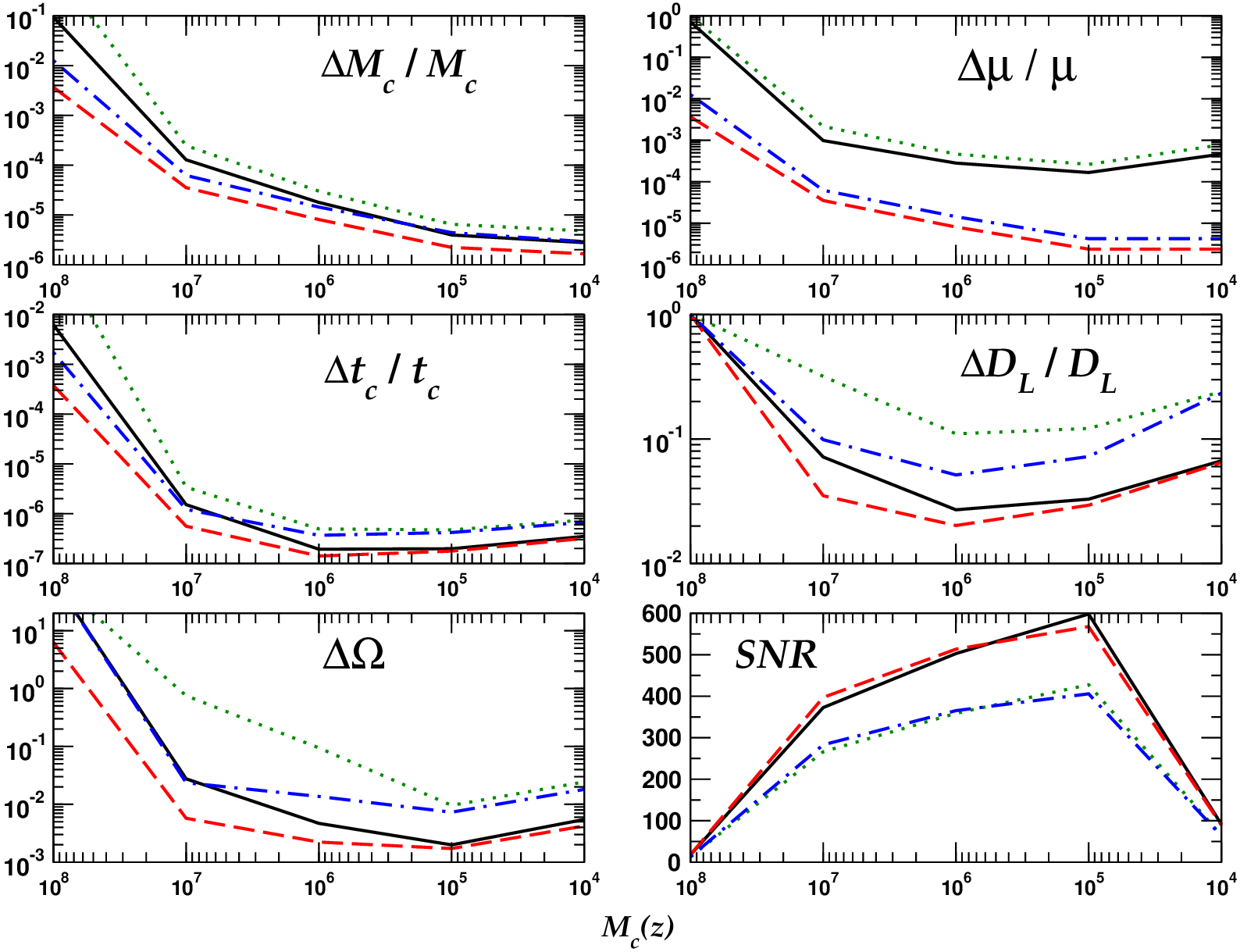}}
\vspace{5mm}
\caption{A comparison of parameter extraction for equal mass binaries using restricted one channel (dotted lines), restricted two channel (solid line) and higher harmonic corrected one channel (dash-dot line) and two channel (dashed line) waveforms, as a function of redshifted chirp mass and at a distance of 10 Gpc.  The values quoted are median values from a $2\times10^4$ point Monte Carlo simulation.}
\label{fig:fem}
\end{center}
\end{figure}
\section{A One versus Two Channel LISA, with higher harmonic corrections.}\label{sec:ovt}
The final issue we look at is the advantage of using HHC waveforms if there is a problem with LISA and we lose two detector links.  While the goal is always to have six links working, the question stands : can we still do astronomy with a broken detector?  In the following text we will refer to one channel restricted results as $X_0$, two channel restricted as $AE_0$, one channel HHC as $X_2$ and two channel HHC results as $AE_2$.  There are two results we can present here, before making distinctions between equal and unequal masses.  The first is that a full LISA is always going to perform better, and the second is that there is in general an improvement in SNR of $\sim\sqrt{2}$ in going from a one to a two channel LISA regardless of the waveform type.

\begin{figure}[t]
\vspace{0.25 in}
\begin{center}
\centerline{\epsfxsize=14cm \epsfysize=10cm \epsfbox{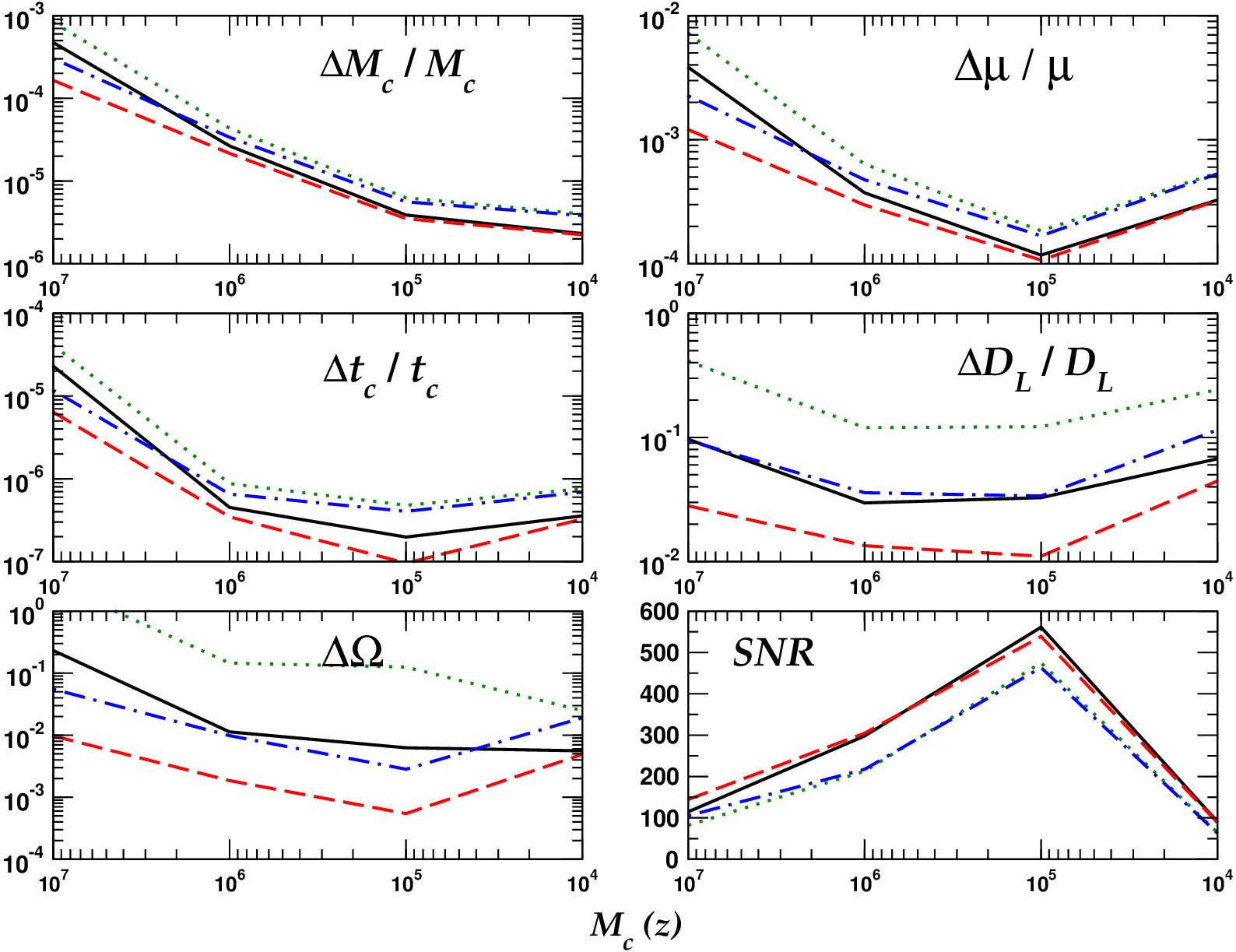}}
\vspace{5mm}
\caption{A comparison of parameter extraction for binaries with a mass ratio of 10 using restricted one channel (dotted lines), restricted two channel (solid line) and higher harmonic corrected one channel (dash-dot line) and two channel (dashed line) waveforms, as a function of redshifted chirp mass and at a distance of 10 Gpc.  The values quoted are median values from a $2\times10^4$ point Monte Carlo simulation.}
\label{fig:fum}
\end{center}
\end{figure}

\subsection{Equal Mass Binaries.}
In Fig~(\ref{fig:fem}), we repeat the results of Fig~(\ref{fig:em1}), this time with the inclusion of parameter estimation just using one LISA channel.  We can see that in all cases, losing one LISA channel and just having the $X_0$ output would seriously hamper our ability to carry out parameter estimation for all parameters.  For the $M_c=10^8\,M_{\odot}$ case, while the signal would be just about detectable over a galaxy with a median SNR of 12.6, we could say nothing about $\{M_{c},\mu,\Delta\Omega,D_{L}\}$ and would have an enormous error in time to coalescence prediction.  As we go to the $M_c=10^7\,M_{\odot}$ case, the gap between the $X_0$ and $AE_0$ results have narrowed for $\{M_{c},\mu,t_c\}$ with a factor of $\sim$ 2 in error.  However for sky resolution and luminosity distance there is a loss of information of factors of 27 and 4.5 respectively.  This is again due to the fact that in one channel, we would only measure one phase and one amplitude, giving two observables for the six unknowns $\{\theta,\phi, \psi, \iota, \varphi_c, \ln(D_L)\}$.  

With the introduction of the HHC waveforms, it is clear, as we can see from the plots, that for a chirp mass of $\gtrsim 2\times10^6\,M_{\odot}$ we can do as well, if not better, using the $X_2$ output as we can using the $AE_0$ channels.  Again at $M_c=10^8\,M_{\odot}$, the sky position and luminosity distance are unresolvable using the $X_2$ channel only.  However, for chirp mass, reduced mass and coalescence time we actually improve the error estimate by factors of almost 10, 57 and 4 respectively over the $AE_0$ channels.  At around $M_c=10^7\,M_{\odot}$ we see the real power of including the HHCs.  Except for luminosity distance, where there is a slight increase in error, the performance of the $X_{2}$ channel equals or exceeds the $AE_0$ channels.  In fact, we see improvement factors of 2 in chirp mass and 16 in reduced mass when we use just the $X_2$ channel.  We can justify this increase in performance by examining the observable count.  We stated earlier that the $AE_0$ combination measures two phases and two amplitudes, giving four observables for six unknowns.  Once we introduce the HHCs in the equal mass case,  we now have three amplitudes and three phases, giving us the six observables we need for the six unknowns.

\subsection{Unequal Mass Binaries.}
We present the same results for the unequal mass case in Fig~(\ref{fig:fum}).  As in the equal mass case, we see some similar patterns.  In all cases an $X_0$ channel alone will seriously effect our ability to do GW astronomy, with the main effect coming in sky resolution and luminosity distance estimations.  We can see that the errors for these two parameters changes by at least an order of magnitude, while there is a loss of almost an order of magnitude for the mass parameters.  For the $X_2$ channel case, we again see that at $M_c \gtrsim 2\times10^6\,M_{\odot}$ the one channel with HHCs does as well if not better than the $AE_0$ channels.  As is expected, the improvement in the masses is not as pronounced in the unequal mass case, but we still see improvements at the highest mass.  The major result in this case is that an $X_2$ channel will outperform the $AE_0$ cases down to a redshifted chirp mass of $\sim10^5\,M_{\odot}$.

It is interesting to examine the effects of the HHCs in the one detector case in a deeper manner.  To do so, we will focus on the the case of a system with a redshifted chirp mass of $10^7\,M_{\odot}$ and a mass ratio of 10.  As usual, the source is at 10 Gpc and is coalescing just inside the observation time.  The first thing we can look at is the observables count.  In this case, the extra harmonics allow us to measure six phases and six amplitudes giving us twelve observables for six unknowns.  In fact, it is now clear that just adding the $H^{(1/2)}_{+,\times}$ terms are enough to outperform the $AE_0$ channels, as with this model we would have three phases and three amplitudes, giving us the six observables needed for the six unknowns.  We should note however, that for the equal mass system, we would need to go to higher harmonic orders due to the null harmonic terms.

\begin{table}
\begin{tabular}{c|cccccccccccc}\hline\hline
 & $\ln(M_{c})$ & $\ln(\mu) $ & $\ln(t_c)$ & $\cos\theta$ & $\phi$ & $\cos\iota$ & $\ln(D_L)$  &$\psi$ & $\varphi_c$\\
\hline\hline
$\Delta X_{0}$ & $8.3\times10^{-4}$ & $7.26\times10^{-3}$ & $4.1\times10^{-5}$ & 0.269 & 2.089 & 0.118 & 0.412 & 0.429 & 2.06 \\
$\Delta X_{2}$ & $3.2\times10^{-4}$& $2.23\times10^{-3}$ & $1.2\times10^{-5}$ & 0.029 & 0.315 & 0.078 & 0.095 & 0.077 & 0.81\\
$\Delta AE_{0}$ & $4.73\times10^{-4}$ & $3.83\times10^{-3}$ & $2.3\times10^{-5}$ & 0.043 & 0.06 & 0.048 & 0.12 & 0.11 & 0.3\\
$\Delta AE_{2}$ & $1.65\times10^{-4}$ & $1.12\times10^{-3}$ & $6.1\times10^{-6}$ & 0.0203 & 0.037 & 0.009 & 0.028 & 0.029 & 0.169 \\
$X_{0}/X_{2}$ & 2.59 & 3.26 & 3.3 & 9.44 & 6.63 & 1.51 & 4.35 & 5.57 & 2.54 \\
$AE_{0}/AE_{2}$ & 2.89 & 3.42 & 3.8 & 2.1 & 1.62 & 5.33 & 4.3 & 3.8 & 1.78\\
$X_{0}/AE_{0}$ & 1.76 & 1.89 & 1.74 & 6.26 & 34.8 & 2.46 & 3.4 & 3.9 & 6.85\\
$X_{2}/AE_{2}$ & 1.8 & 1.99 & 2.0 & 1.4 & 8.5 & 8.67 & 3.34 & 2.6 & 4.79\\
\hline\hline
\end{tabular}
%\end{indented}
\caption{Median errors for a source with a redshifted chirpmass of $10^7\,M_{\odot}$ and a mass ratio of 10 at a distance of 10 Gpc. For each parameter we show the one-channel LISA correlations without HHCs, $X_0$, one-channel LISA correlations with HHCs, $X_2$, two-channel LISA correlations without HHCs, $AE_0$, and two-channel LISA correlations with HHCs, $AE_2$, errors. The last four rows show the ratios of the median error values for various channel combinations.  The errors for the angular variables are in radians.}
\label{tab:errs}
\end{table}

\begin{figure}
\begin{center}
\centerline{\epsfxsize=12cm \epsfysize=8cm \epsfbox{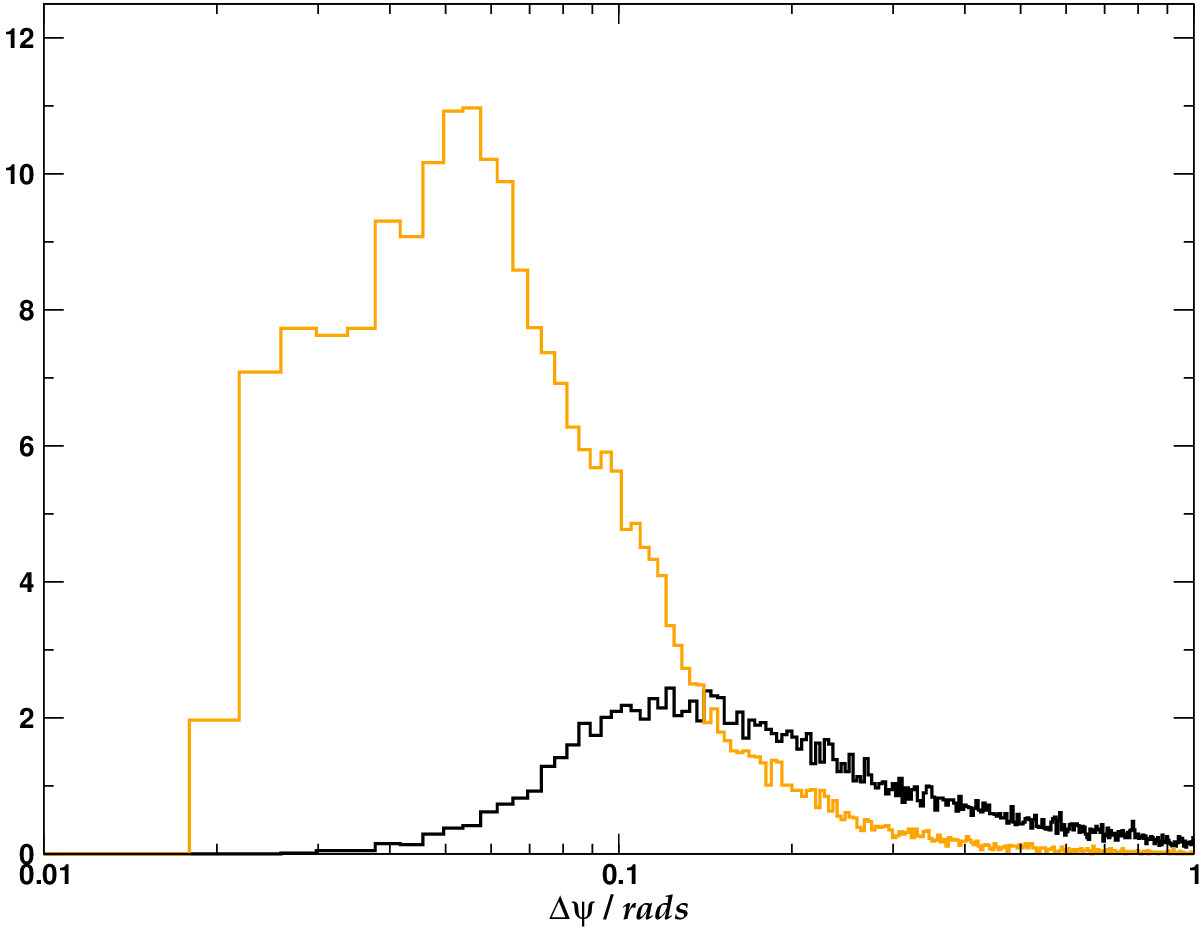}}
\vspace{5mm}
\caption{Histograms for the error in the polarization angle $\psi$ for restricted (dark) and corrected (light) waveforms using a single LISA detector, for systems with a redshifted chirpmass of $10^7\,M_{\odot}$, a mass ratio of 10, at a distance of 10 Gpc.}
\label{fig:psierror}
\end{center}
\end{figure}

In Table~\ref{tab:errs} we present the median errors for a Monte Carlo simulation of $2\times10^4$ systems.  We can see from having the extra observables, how the $X_2$ channel fares against the $X_0$ channel, and more importantly, the $AE_0$ channels.  As long as we include the HHCs, a one detector LISA would still measure $\{M_c,\mu,t_c,\cos\theta,\psi,\ln(D_L)\}$ factors of (1.5,1.7,1.9,1.5,1.3,1.4) times better than a two detector LISA without.  We should point out here that one of the most important improvement comes in the measurement of the polarization angle $\psi$.  In Fig~(\ref{fig:psierror}) we plot histograms for the error in the estimation of $\psi$ with and without HHCs.  We can see that having the extra HHCs allows us to make an instantaneous measurement of $\psi$ which is important firstly, in determining the binaries principle polarization axis around the line of sight, and secondly, in the overall amplitude of the detector response.

\begin{figure}
\begin{center}
\centerline{\epsfxsize=12cm \epsfysize=8cm \epsfbox{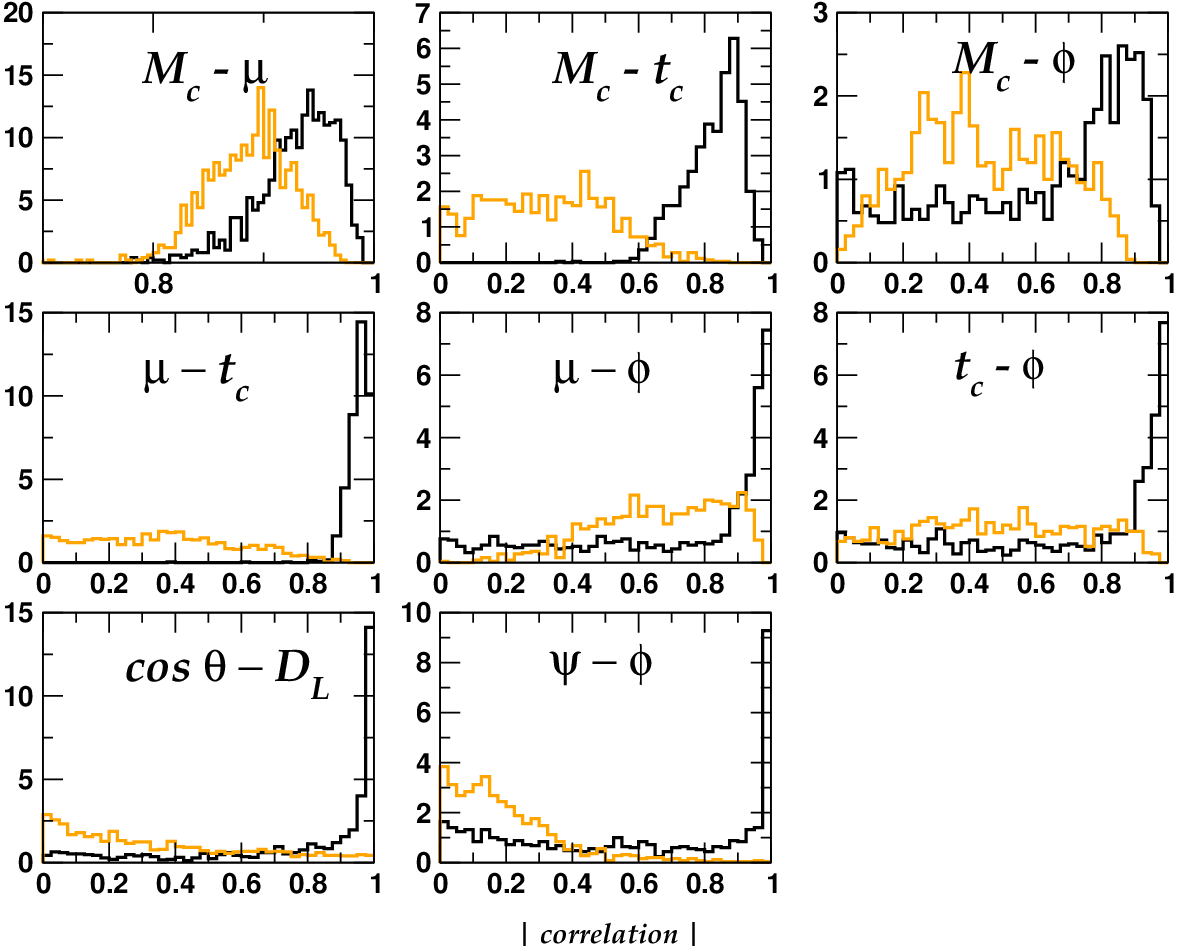}}
\vspace{5mm}
\caption{An examination of parameter correlations for the case of a single LISA detector for systems with a redshifted chirpmass of $10^7\,M_{\odot}$, a mass ratio of 10, at a distance of 10 Gpc. The dark lines denote the restricted waveforms, while the light lines signify the corrected waveforms.  }
\label{fig:1linkcorr}
\end{center}
\end{figure}

The final thing we will investigate is the correlations between the parameters.  It is interesting to trace the evolution in correlations to see the effects of the inclusion of HHCs and then the inclusion of a second channel.  In Fig~(\ref{fig:1linkcorr}), we display some of the main correlation breaking with a single detector.   In Table~\ref{tab:cov} we tabulate the correlations in both the one-detector case with and without HHCs, and in the two detector case with HHCs.  As displayed in previous sections, the parameters $\{M_c,\mu,t_c , \phi\}$ are all highly correlated with each other.  However, the introduction of the HHCs again cause a decoupling between these parameters.  As well as that we also notice a decoupling between $\cos\theta$ and $D_L$ and importantly, between $\phi$ and $\psi$.  It would seem that the decoupling of $\phi$ with four other parameters is responsible for the improvement in sky resolution at high masses.

When we move to the two detector case with HHCs, we can see that the extra channel now serves to refine the parameter estimation.  This is due to a slight further decrease in some of the parameter correlations.  While we have not shown the results here, the introduction of a second channel only causes a slight decrease in the breaking of correlations.  It is clear that it is the HHCs that are having the main effect.

\begin{table}
\begin{tabular}{c|c|cccccccccccc}\hline\hline
 &  & $\ln(\mu) $ & $\ln(t_c)$ & $\cos\theta$ & $\phi$ & $\psi$ &$\cos\iota$ & $\ln(D_L)$ & $\varphi_c$\\
\hline\hline
$\ln(M_{c})$ & $X_0$  & 0.94 & 0.84 & 0.22 & 0.69 & 0.17 & 0.19 & 0.18 & 0.17 \\
& $X_2$  & 0.89 & 0.32 & 0.10 & 0.42 & 0.16 & 0.22 & 0.18 & 0.15 \\
& $AE_2$  & 0.87 & 0.24 & 0.07 & 0.40 & 0.08 & 0.09 & 0.09 & 0.06 \\ \hline
$\ln(\mu)$ & $X_0$ & * & 0.98 & 0.29 & 0.85 & 0.23 & 0.24 & 0.23 & 0.21\\
& $X_2$ & * & 0.36 & 0.14 & 0.68 & 0.18 & 0.28 & 0.21 & 0.18 \\
& $AE_2$ & * & 0.22 & 0.11 & 0.68 & 0.10 & 0.13 & 0.12 & 0.08 \\ \hline
$\ln(t_{c})$ & $X_0$& * & * & 0.25 & 0.84 & 0.22 & 0.22 & 0.20 & 0.18\\
& $X_2$ & * & * & 0.17 & 0.48 & 0.23 & 0.25 & 0.24 & 0.28 \\
& $AE_2$ & * & * & 0.16 & 0.54 & 0.16 & 0.13 & 0.17 & 0.15 \\ \hline
$\cos\theta$ & $X_0$ & * & * & * & 0.26 & 0.22 & 0.33 & 0.93 & 0.40 \\
& $X_2$ & * & * & * & 0.15 & 0.06 & 0.06 & 0.27 & 0.07 \\
& $AE_2$ & * & * & * & 0.12 & 0.13 & 0.19 & 0.42 & 0.19 \\ \hline
$\phi$ & $X_0$ & * & * & * & * & 0.60 & 0.28 & 0.21 & 0.30 \\
& $X_2$ & * & * & * & * & 0.16 & 0.12 & 0.11 & 0.10 \\
& $AE_2$ & * & * & * & * & 0.24 & 0.11 & 0.10 & 0.09 \\ \hline
$\psi$ & $X_0$ & * & * & * & * & * & 0.36 & 0.28 & 0.50\\
& $X_2$ & * & * & * & * & * & 0.43 & 0.57 & 0.20 \\
& $AE_2$ & * & * & * & * & * & 0.28 & 0.45 & 0.09 \\ \hline
$\cos\iota$ & $X_0$ & * & * & * & * & * & * & 0.42 & 0.27\\
& $X_2$ & * & * & * & * & * & * & 0.57 & 0.20 \\
& $AE_2$ & * & * & * & * & * & * & 0.45 & 0.09 \\ \hline
$\ln(D_L)$ & $X_0$ & * & * & * & * & * & * & * & 0.47\\
& $X_2$ & * & * & * & * & * & * & * & 0.63 \\
& $AE_2$ & * & * & * & * & * & * & * & 0.58 \\
\hline\hline
\end{tabular}
%\end{indented}
\caption{Absolute median values of the correlation matrix for a source with a redshifted chirpmass of $10^7\,M_{\odot}$ and a mass ratio of 10 at a distance of 10 Gpc.  For each parameter we show the one-channel LISA correlations without HHCs, $X_0$, one-channel LISA correlations with HHCs, $X_2$, and two-channel LISA correlations with HHCs, $AE_2$.}
\label{tab:cov}
\end{table}

\section{Conclusion}
In this work we have looked at the effect of including higher harmonic corrections to the restricted PN waveforms in the LISA context.  By carrying out an extensive Monte Carlo simulation for various redshifted chirp masses and mass ratios, we have tried to cover as many scenarios as possible.  As was already known, the corrected waveforms bring previously invisible sources into the LISA bandwidth.  We have shown however, that in general while we will be able to detect these sources, we will not be able to say anything useful for any electromagnetic follow-up.  For equal mass binaries, due to a combination of correlation breaking and the reduced mass becoming perfectly correlated with the chirp mass, there is a significant improvement in mass estimation using the corrected waveforms.  We found that while there are systems where improvements in sky resolution and luminosity distance are huge, the median improvements are not as dramatic as some individual sources.  The same is true in the unequal mass case, although here there is an order of magnitude improvement in sky resolution.  We also showed that the overall effect of the HHCs is providing enough observables to account for the number of unknown parameters.

We finally looked at the more interesting case where the LISA output was reduced to one channel.  We showed here that for supermassive and some massive systems, we can actually do better with a one channel LISA with HHCs than we can with a full LISA with no harmonic corrections.  As the HHCs again provide enough observables to solve for the unknown parameters, this allows us to improve mass measurement, sky resolution and make instantaneous measurements of the polarization $\psi$.  While a second channel always improves the parameter extraction, it is clear that the HHCs are the main source of correlation breaking and improvement in parameter estimation.  This is an important result as it shows that while we would always like a full LISA, not all would be lost if we were forced to work with a single channel detector.

\section{Acknowledgments}
EKP would like to thank the DLR (Deutsches Zentrum f\"ur Luft- und Raumfart) for support during this work.  EKP would also like to thank Leor Barack and Cole Miller for stimulating conversations.

\end{document}